%% file: advanced_lithography_2008.tex
\documentclass[a4paper]{spie}  
\usepackage[]{graphicx}
\usepackage{amssymb,amsmath,psfrag,url}
\include{MatheDef}

\title{A Rigorous Finite-Element Domain Decomposition Method for Electromagnetic Near Field Simulations}

\author{
Lin Zschiedrich\supit{\,ab},
Sven Burger\supit{\,ab}, 
Achim Sch\"adle\supit{\,a},
Frank Schmidt\supit{\,ab}, 
\skiplinehalf
\supit{a}
Zuse Institute Berlin,
Takustra{\ss}e 7,
D\,--\,14\,195 Berlin,
Germany
\smallskip\\
\supit{b}
JCMwave GmbH,
Haarer Stra{\ss}e 14a,
D\,--\,85\,640 Putzbrunn, 
Germany
}

\authorinfo{
Corresponding author: L. Zschiedrich\\
URL: http://www.zib.de/Numerik/NanoOptics/\\
Email: zschiedrich@zib.de
}

\begin{document} 
\maketitle 
\noindent
Copyright 2008  Society of Photo-Optical Instrumentation Engineers.\\
This paper will be published in Proc.~SPIE Vol. {\bf 6924}
(2008),  
({\it Optical Microlithography XXI, H.\,J.~Levinson, M.\,V.~Dusa, Eds.})
and is made available 
as an electronic preprint with permission of SPIE. 
One print or electronic copy may be made for personal use only. 
Systematic or multiple reproduction, distribution to multiple 
locations via electronic or other means, duplication of any 
material in this paper for a fee or for commercial purposes, 
or modification of the content of the paper are prohibited.
\begin{abstract}
Rigorous computer simulations of propagating electromagnetic fields have
become an important tool for optical metrology and design of nanostructured
optical components. A vectorial finite element method (FEM) is a good choice
for an accurate modeling of complicated geometrical features. However, from a
numerical point of view solving the arising system of linear equations is very
demanding even for medium sized 3D domains. In numerics, a domain decomposition
method is a commonly used strategy to overcome this problem. Within this
approach the overall computational domain is split up into smaller domains
and interface conditions are used to assure continuity of the electromagnetic
field. Unfortunately, standard implementations of the domain decomposition
method as developed for electrostatic problems are not appropriate for wave propagation
problems. In an earlier paper we therefore proposed a domain decomposition method adapted to
electromagnetic field wave propagation problems. In this paper we apply this method to 3D mask simulation.
  
\end{abstract}



\section{Introduction}
\begin{figure}
 \psfrag{e, E, H}{$\epsilon$, $\MyField{E}$, $\MyField{H}$}
 \psfrag{e1, E, H}{$\epsilon_1$, $\MyField{E}_1$, $\MyField{H}_1$}
 \psfrag{e2, E, H}{$\epsilon_2$, $\MyField{E}_2$, $\MyField{H}_2$}
 \psfrag{e3, E, H}{$\epsilon_3$, $\MyField{E}_3$, $\MyField{H}_3$}
 \psfrag{Einc}{$\MyField{E}_{\mathrm{inc}}$}
  \begin{center}
      \includegraphics[width=17cm]{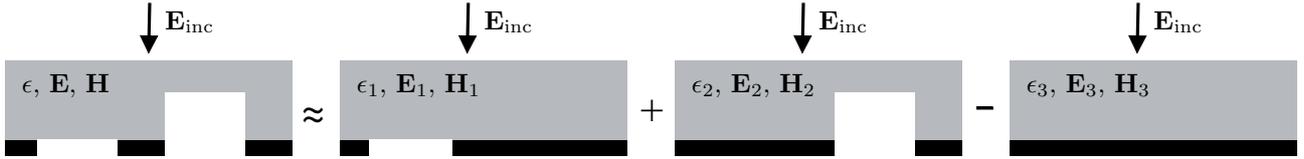}
  \end{center}
  \caption{
\label{Fig:NeureutherDD}Domain decomposition method by Adam and Neureuther (not applied in this paper). One uses a superposition
of the fields $\MyField{E}_1$, $\MyField{E}_2$, and $\MyField{E}_3$ to approximate the field
$\MyField{E}.$ Although this approach yields satisfactory results for many applications in mask design it introduces errors inversely proportional to the distance of the features.
}
\end{figure}

\begin{figure}
 \psfrag{e, E}{$\epsilon$, $\MyField{E}$}
 \psfrag{e1, E1}{$\epsilon_1$, $\MyField{E}_1^{(k)}$}
 \psfrag{e2, E2}{$\epsilon_2$, $\MyField{E}_2^{(k)}$}
 \psfrag{Einc}{$\MyField{E}_{\mathrm{inc}}$}
 \psfrag{E1sc}{$\MyField{E1}_{1, \mathrm{sc}}^{(k)}$}
 \psfrag{E2sc}{$\MyField{E2}_{2, \mathrm{sc}}^{(k)}$}
 \psfrag{Rand}{$\Gamma$}
  \begin{center}
      \includegraphics[width=12cm]{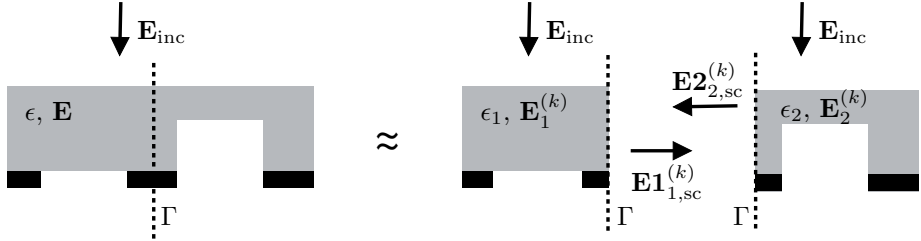}
  \end{center}
  \caption{Domain decomposition method as discussed in this paper (horizontal
    setting). For each sub-domain transparent boundary conditions are
    imposed. Starting with the first sub-domain the field $\MyField{E}_1^{(0)}$ and the scattered field $\MyField{E}_{1, \mathrm{sc}}^{(0)}$ to the incoming wave $\MyField{E}_{\mathrm{inc}}$ are computed. Afterwards the second domain is updated, where the incoming field is now the sum of the originally incident field $\MyField{E}_{\mathrm{inc}}$ and the scattered field $\MyField{E}_{1, \mathrm{sc}}^{(0)}.$ This process is iteratively repeated. Observe that the zeroth iterate is exactly the method by Adam and Neureuther. 
\label{Fig:HorizontalDD}
}
\end{figure}

\begin{figure}
 \psfrag{e, E}{$\epsilon$, $\MyField{E}$, $\MyField{H}$}
 \psfrag{e1, E1}{$\epsilon_1$, $\MyField{E}_1$, $\MyField{H}_1$}
 \psfrag{e2, E2}{$\epsilon_2$, $\MyField{E}_2$, $\MyField{H}_2$}
 \psfrag{Einc}{$\MyField{E}_{\mathrm{inc}}$}
 \psfrag{E1sc}{$\MyField{E1}_{1, \mathrm{sc}}^{(k)}$}
 \psfrag{E2sc}{$\MyField{E2}_{2, \mathrm{sc}}^{(k)}$}
 \psfrag{Rand}{$\Gamma$}
  \begin{center}
      \includegraphics[width=13cm]{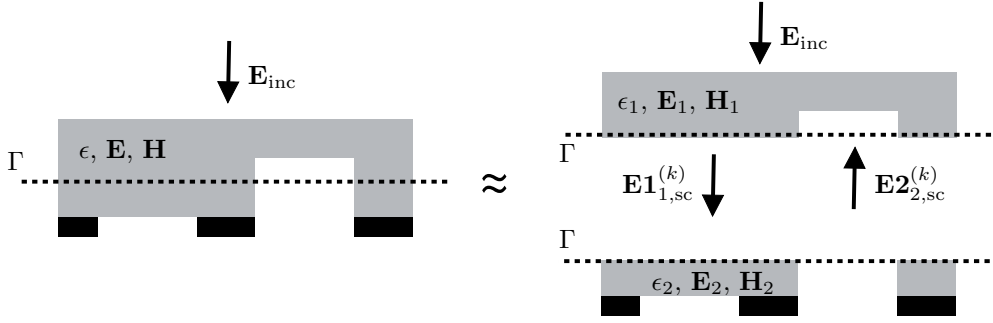}
  \end{center}
  \caption{Domain decomposition method as discussed in this paper (vertical setting). The notation is given in Figure~\ref{Fig:HorizontalDD}. 
\label{Fig:VerticalDD}
}
\end{figure}

This paper adresses on the rigorous simulation of the
electromagnetic field within a cutout of a microlithography mask or semiconductor wafer. It might be
confusing that the notion ``domain decomposition method'' appears in the
numerical literature about Maxwell's equations and in the mask simulation
community with -- at a first glance -- different meanings. In the mask
simulation community, Adam and Neureuther introduced the domain decomposition
method in 2001\cite{Adam:01}. As an example, in the approach by Adam and
Neureuther a phase shift line mask is split up into two isolated lines, c.f. Figure~\ref{Fig:NeureutherDD}. 
The transmission through each line is computed separately. Afterwards the computed fields are merged in an appropriate way. Diffraction effects
of the double slit are neglected in a first step but can also be approximated
by adding further cross talk terms. 
In the numerics community the domain
decomposition method denotes a strategy for solving large scale partial differential
equation problems and goes back to Schwarz more than 100 years ago
\cite{Quarteroni:99}. This approach shares the key idea by Adam and Neureuther
to split the computational problem into smaller, more feasible
subproblems. But in contrast to the method by Adam and Neureuther the
synthesis step (merging the fields on the sub-domains) relies on an iterative
numerical method and not on physical intuition. This way the method remains
rigorous: Starting from a rigorous discretization of Maxwell's equations in
the entire domain one
gets a large scale linear system of equations. The aim is to iteratively solve
this large scale linear system. As long as the iterative process converges this
yields the rigorous solution. Although the domain decomposition method is well
established and well analyzed for electrostatic and eddy current problems the
application of  domain decomposition techniques to high frequency Maxwell's
equations is still an active research area in the numerics
community. Especially, for wave propagation problems it is necessary to impose
proper radiation boundary conditions at the coupling boundary of the adjacent
sub-domains \cite{Zschiedrich:05, Schaedle:07}. We will detail this issue in
the next section. Figure~\ref{Fig:HorizontalDD} and
Figure~\ref{Fig:VerticalDD} give a rough sketch of the algorithm. In both
examples the computational domain is split into two isolated structures.
Starting with the first sub-domain the field $\MyField{E}_1^{(0)}$ and the
scattered field $\MyField{E}_{1, \mathrm{sc}}^{(0)}$ to the incoming wave
$\MyField{E}_{\mathrm{inc}}$ are computed. Afterwards the second domain is
updated, where the incoming field is now the sum of the originally incident
field $\MyField{E}_{\mathrm{inc}}$ and the scattered field $\MyField{E}_{1,
  \mathrm{sc}}^{(0)}.$ This process is iteratively repeated. This approach
resembles the multiple scattering method~\cite{Martin:06a}. Also observe that
the zeroth iterate is exactly the method by Adam and Neureuther.  \\
The paper is structured as follows. Section~\ref{sec:modeling} is devoted to
the modeling of Maxwell's equations and of the domain decomposition
iteration. We further summarize various numerical issues such as the definition of
transparent boundary conditions and casting Maxwell's equations into weak form
needed for the finite element method. The derivation of a weak form for Maxwell's equations on structured unbounded domains
was already discussed in our papers~\cite{Zschiedrich2005a, Zschiedrich2006a}, but also see the recent paper by Wei et al.\cite{Wei:07} In section~\ref{sec:numerical_example}
we apply the domain decomposition method to a medium sized cutout of a
mask. The paper ends with some concluding remarks.  

\section{Modeling}
\label{sec:modeling}
\subsection{Maxwell's equations}
Starting from Maxwell's equations in a medium without sources and free currents and assuming time-harmonic dependence with angular frequency $\omega>0$ the electric and magnetic fields
\[
\VField{E}(x, y, z, t)  = \widetilde{\VField{E}}(x, y, z)e^{-i\omega \cdot t}, \;
\VField{H}(x, y, z, t) = \widetilde{\VField{H}}(x, y, z)e^{-i\omega \cdot t}, \;
\]
must satisfy 
\begin{eqnarray*}
\nabla \times \widetilde{\VField{E}} & = & i\omega \mu \widetilde{\VField{H}}, \quad
\nabla \cdot \epsilon \widetilde{\VField{E}} = 0, \\
\nabla \times \widetilde{\VField{H}} & = & -i\omega \epsilon \widetilde{\VField{E}}, \quad 
\nabla \cdot \mu \widetilde{\VField{H}} = 0. \\
\end{eqnarray*}
Here $\epsilon$ denotes the permittivity tensor and $\mu$ denotes the permeability tensor of the materials. In the following we drop the wiggles, so that $\widetilde{\VField{E}} \rightarrow \VField{E}$, $\widetilde{\VField{H}} \rightarrow \VField{H}$. From the equations above we then may derive (by direct substitution) the second order equation for the electric field
\begin{eqnarray*}
\nabla \times \mu^{-1} \nabla \times \VField{E} - \omega^2 \epsilon \VField{E} & = & 0, \\
\nabla \cdot \epsilon \VField{E} & = & 0.
\end{eqnarray*}  
A similar equation holds true for the magnetic field - one only need to replace $\VField{E}$ by $\VField{H}$ and interchange $\epsilon$ and $\mu$. Observe that any solution to the first equation also meets the divergence condition (second equation). We therefore drop the second equation in the following.\\
For the sake of a simpler and more compact notation we rewrite these equations in differential form,
\begin{eqnarray}
\label{thefieldmaxequationsdf}
d_{1} \mu^{-1}  d_{1} \VField{e} - \omega^2 \epsilon \VField{e} & = & 0.
\end{eqnarray}  
A reader not familiar with this calculus may replace the exterior derivatives $d_{0}$, $d_{1}$, $d_{2}$ with classical differential operators, $d_{0} \rightarrow \nabla$, $d_{1} \rightarrow \nabla \times $ and $d_{2} \rightarrow \nabla \cdot$. Here, the electric field appears as a differential 1-form, $e = e_{x}dx+e_{y}dy+e_{z}dz,$ whereas the material tensors act -- from a more mathematical point of view -- as operators
\begin{eqnarray*}
\epsilon,\, \mu \; : \; \Alt^{1} \rightarrow \Alt^{2}.
\end{eqnarray*}   
In the following we drop the sub-indices for the exterior derivatives $d_{0}$, $d_{1}$, $d_{2}.$
\subsection{Scattering off an isolated structure} 
\begin{figure}
 \psfrag{x}{$x$}
 \psfrag{y}{$y$}
 \psfrag{z}{$z$}
 \psfrag{Einc}{$\MyField{E}_{\mathrm{inc}}$}
 \psfrag{Esc}{$\MyField{E}_{\mathrm{sc}}$}
 \psfrag{Rand}{$\Gamma$}
  \begin{center}
      \includegraphics[width=10cm]{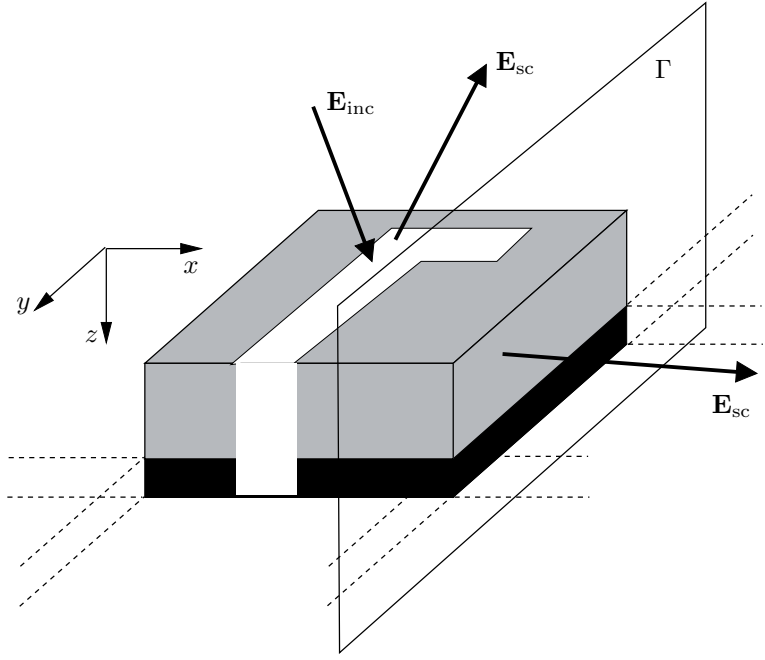}
  \end{center}
  \caption{Sketch of a 3D computational domain. The structure is embedded into
    a multi-layer stack (mask or wafer blank). The incoming field
    $\MyField{E}_{\mathrm{inc}}$ is a solution to Maxwell's equation in the
    layered media. Typically a plane wave traveling in the $z$-direction 
    and scattered at the multi-layer stack is used. Periodic problems can also
    be dealt with but are not specially detailed in this paper. The plane
    $\Gamma$ is used in the domain decomposition process. Within this context
    the sketched domain is only a sub-domain of the entire domain. $\Gamma$ separates two regions which are merged in the domain decomposition iteration.
\label{Fig:IsolatedBlock}
}
\end{figure}

We now deal with light scattering off an isolated structure as depicted in Figure~\ref{Fig:IsolatedBlock}. 
The block  $\Omega = [-a,\, a] \times [-b,\,  b] \times [-c,\,  c]$ is the
computational domain. For simplicity we assume that the computational domain
is aligned along the coordinate axes and is embedded
into a layered media. The role of the hyperplane $\Gamma$ will be explained later. The incident field $\MyField{E}_{\mathrm{inc}}$ is a solution to Maxwell's equations in the layered media. Typically  $\MyField{E}_{\mathrm{inc}}$ is a plane wave traveling in the $z$-direction scattered at the multi-layer stack. Outside the computational domain the total field is a sum of the incident field and the scattered field $\MyField{E}_{\mathrm{sc}}$, which also is a solution to Maxwell's equations in the layered media outside the computational domain. The interior and exterior fields are linked by the following coupled system, 
\begin{subequations}
\label{strongcoupledsystem}
\begin{eqnarray}
d \mu^{-1}  d \VField{e}_{\mathrm{int}} - \omega^2 \epsilon \VField{e}_{\mathrm{int}} & = & 0, \quad \mbox{on}\; \Omega \\
d \mu^{-1}  d \VField{e}_{\mathrm{sc}} - \omega^2 \epsilon \VField{e}_{\mathrm{sc}} & = & 0, \quad \mbox{on}\; \rnum^3 \setminus \Omega \\
\label{strongcoupledsystemDirichlet}
\VField{e}_{\mathrm{int}}- \VField{e}_{\mathrm{sc}} & = & \VField{e}_{\mathrm{inc}},  \quad \mbox{on}\; \partial  \Omega \\
\label{strongcoupledsystemNeumann}
 \mu^{-1}_{\mathrm{int}} d \VField{e}_{\mathrm{int}}-  \mu^{-1}_{\mathrm{ext}} d \VField{e}_{\mathrm{sc}} & = &  \mu^{-1}_{\mathrm{ext}} d \VField{e}_{\mathrm{inc}},  \quad \mbox{on}\; \partial  \Omega. 
\end{eqnarray}
\end{subequations}
The first two equations are Maxwell's equations for the interior and the scattered field. Equation~\eqref{strongcoupledsystemDirichlet} and equation~\eqref{strongcoupledsystemNeumann} assure the tangential continuity of the electric field and the magnetic field respectively. 
\subsection{Domain decomposition iteration}
\begin{figure}
 \psfrag{O1}{$\Omega_1$}
 \psfrag{O2}{$\Omega_2$}
 \psfrag{O3}{$\Omega_3$}
 \psfrag{Esc3}{$\MyField{E}_{\mathrm{sc}, 3}$}
 \psfrag{G12}{$\Gamma_{1, 2}$}
 \psfrag{G23}{$\Gamma_{2, 3}$}
  \begin{center}
      \includegraphics[width=10cm]{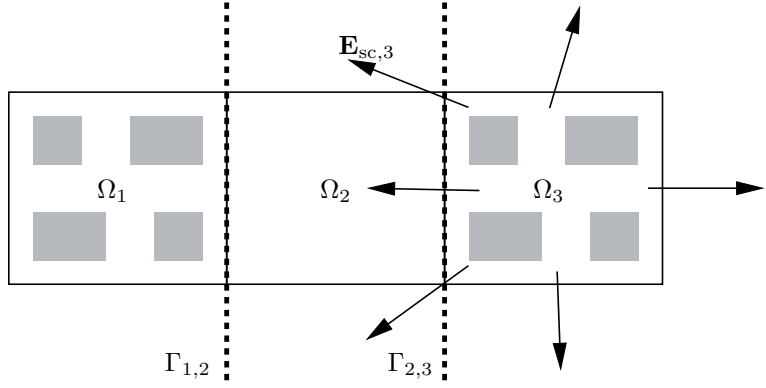}
  \end{center}
  \caption{Domain decomposition of the entire domain in three sub-domains
    aligned in a line. As an instructive example, the middle domain does not
    contain a scatterer. The field $\MyField{E}_{\mathrm{sc}, 3}$ scattered
    from domain $\Omega_3$ enters the middle domain across the infinite interface
    $\Gamma_{2, 3}$. In the domain decomposition process the field travels
    through the middle domain and hits domain $\Omega_1$ across the interface
    $\Gamma_{1, 2}.$ 
\label{Fig:SimpleChain}
}
\end{figure}

In the domain decomposition method the entire computational domain is split
into $n$ smaller boxes $\Omega_i.$  Initially, one starts with one sub-domain
and solves the coupled interior-exterior
system~\eqref{strongcoupledsystem}. When updating the subsequent domains we
account for the already computed scattered field of the adjacent domains.\\
Let us first regard the instructive example as in Figure~\ref{Fig:SimpleChain}. There the
entire domain is split into three sub-domains aligned in a linear chain. Assume
that initially only the scattered field on domain $\Omega_3$ is computed. Next
the middle domain $\Omega_2$ is updated. We solve
system~\eqref{strongcoupledsystem} where
the incident field is now the sum of the original light source field and the
scattered field $\MyField{E}_{\mathrm{sc}, 3}$ from domain $\Omega_3.$ Since
the domain $\Omega_2$ does not contain any scatterer the scattered field
$\MyField{E}_{\mathrm{sc}, 2}$  from this domain is zero. To update domain
$\Omega_1$ now we have to gather all waves which travel through interface
$\Gamma_{1, 2}.$ Obviously, this field consists of the field scattered from domain
$\Omega_3$ travelled through $\Omega_2$ and the original source light from the
illumination system. The key issue here is, that it is not sufficient to
compute the scattered field for $\Omega_2$ (which is zero here). We also need to
compute the field travelling from one parallel interface to the other.    
\\
We now deal with light scattering off an isolated sub-domain as depicted in
Figure~\ref{Fig:IsolatedBlock}. We introduce the infinite hyperplanes $\Gamma_{i, j}$ which separates the two sub-domains $\Omega_i$ and $\Omega_j$ with normal $n_{i,j}$ directed from $\Omega_i$ to $\Omega_j$, see hyperplane $\Gamma$ in Figure~\ref{Fig:IsolatedBlock}. We define the restriction of the scattered field $\VField{e}_{\mathrm{sc}, i}$ onto the hyperplane 
\[
\VField{e}_{\mathrm{sc}, \Gamma_{i, j}} = \left( \VField{e}_{\mathrm{sc}, i} \right)_{|\Gamma_{i, j}}.    
\]
These data are stored in the domain decomposition iteration, so that in the $k$th iteration data 
$\VField{e}_{\mathrm{sc}, \Gamma_{i, j}}^{(k)}$ are given. For the update step
\[
\VField{e}_{\mathrm{sc}, \Gamma_{i, j}}^{(k)} \longrightarrow \VField{e}_{\mathrm{sc}, \Gamma_{i, j}}^{(k+1)} 
\]
we solve the following system on $\Omega_i$ which accounts for propagation
through the sub-domains as well as scattering within the sub-domains 
\begin{subequations}
\label{strongcoupledsystemupdate}
\begin{eqnarray}
d \mu^{-1}  d \VField{e}_{\mathrm{int}} - \omega^2 \epsilon \VField{e}_{\mathrm{int}} & = & 0, \quad \mbox{on}\; \Omega \\
\label{strongcoupledsystemupdatesc}
d \mu^{-1}  d \VField{e}_{\mathrm{sc}} - \omega^2 \epsilon \VField{e}_{\mathrm{sc}} & = & 0, \quad \mbox{on}\; \rnum^3 \setminus \left ( \Omega \cup \Gamma_{i, \cdot } \right )  \\
\label{strongcoupledsystemDirichletupdate}
\VField{e}_{\mathrm{int}}- \VField{e}_{\mathrm{sc}} & = & \VField{e}_{\mathrm{inc}} + \VField{e}_{\mathrm{sc}, \Gamma_{j, i}}^{(k)},  \quad \mbox{on}\; \partial  \Omega \cap  \Gamma_{i, j} \\
\label{strongcoupledsystemNeumannupdate}
 \mu^{-1}_{\mathrm{int}} d \VField{e}_{\mathrm{int}}- \mu^{-1}_{\mathrm{ext}} d \VField{e}_{\mathrm{sc}} & = & \mu^{-1}_{\mathrm{ext}} d \VField{e}_{\mathrm{inc}} +\mu^{-1}_{\mathrm{ext}} d \VField{e}_{\mathrm{sc}, \Gamma_{j, i}}^{(k)},  \quad \mbox{on}\; \partial  \Omega \cap  \Gamma_{i, j} \\
\label{strongcoupledsystemDirichletupdateHyperplane}
\VField{e}_{\mathrm{sc},-}- \VField{e}_{\mathrm{sc}, +} & = & \VField{e}_{\mathrm{sc}, \Gamma_{j, i}}^{(k)},  \quad \mbox{on}\;    \Gamma_{i, j} \setminus \partial \Omega \\
\label{strongcoupledsystemNeumannupdateHyperplane}
 \mu^{-1}_{\mathrm{ext}} d \VField{e}_{\mathrm{sc}, -}- \mu^{-1}_{\mathrm{ext}} d \VField{e}_{\mathrm{sc}, +} & = &   \mu^{-1}_{\mathrm{ext}} d \VField{e}_{\mathrm{sc}, \Gamma_{j, i}}^{(k)},  \quad \mbox{on}\;  \Gamma_{i, j} \setminus \partial \Omega. 
\end{eqnarray}
\end{subequations}
With equation~\eqref{strongcoupledsystemupdatesc} the so defined scattered field $\VField{e}_{\mathrm{sc}}$ meets Maxwell's equations in the exterior domain but may jump across the hyperplanes $\Gamma_{i, j}$. For each hyperplane $\Gamma_{i, j}$ the jump is defined in equations~\eqref{strongcoupledsystemDirichletupdate}, \eqref{strongcoupledsystemDirichletupdateHyperplane} and is equal to the scattered field of the corresponding adjacent domain. On $\Gamma_{i,j}$ the quantity $\VField{e}_{\mathrm{sc},+}$ is the field limit in the  $n_{i, j}$-direction and  $\VField{e}_{\mathrm{sc},-}$ is the field limit in the opposite direction. As the update step we now define 
\[
\VField{e}_{\mathrm{sc}, \Gamma_{i, j}}^{(k+1)} = \left( \VField{e}_{\mathrm{sc}, +} \right)_{|\Gamma_{i, j}}.
\]
From equations~\eqref{strongcoupledsystemDirichletupdate} and~\eqref{strongcoupledsystemNeumannupdate} one shows that under convergence of the iteration the field continuity across $\Gamma_{i, j} \cap \partial \Omega$ is satisfied. Hence the so merged field satisfies Maxwell's equations on the entire domain.    

\subsection{Weak formulation}
To apply the finite element method we need to transform the linear systems
used in the previous section to a so called weak form. The incorporation of
the various jump conditions in system~\eqref{strongcoupledsystemupdate} is
very technical, but fits seamless into the finite element framework. We
exemplify this only for the simpler system of
equations~\eqref{strongcoupledsystem}. The treatment of the
system~\eqref{strongcoupledsystemupdate} will be commented at the end of
sub-section \ref{sec:pml}. 
In order to derive a weak formulation we define the following function space on the domain $\Omega = \rnum^3$ 
\begin{eqnarray*}
\hcl & = & \left \{\VField{e} \in \Alt^{1} \, | \, (e_{x}, e_{y}, e_{z}) \in (\ltl)^3, \; \nabla \times (e_{x}, e_{y}, e_{z})^{\mathrm T}  \in (\ltl)^3 \right \}. 
\end{eqnarray*}
The weak form of equations~\eqref{thefieldmaxequationsdf} now reads
\begin{eqnarray}
\label{thefieldmaxequationsdfweak}
\int_{\rnum^{3}} \left(  \mu^{-1} d \VField{e} \wedge d \overline{\VField{v}}  - 
\omega^{2} (\epsilon \VField{e}) \wedge \overline{\VField{v}} \right ) & = & 0 
\end{eqnarray}
for all $\VField{v} \in \hcl$ with compact support. 
We now look for a variational formulation, where the incoming field only appears on the right hand side. To do that let us denote the computational domain by $\Omega$ and split Maxwell's equations~\eqref{thefieldmaxequationsdfweak} into an interior and exterior part, 
\begin{eqnarray*}
\int_{\Omega} \left(  \mu^{-1} d \VField{e}_{\mathrm{int}} \wedge d \overline{\VField{v}}  - 
\omega^{2} (\epsilon \VField{e}_{\mathrm{int}}) \wedge \overline{\VField{v}} \right ) 
+
\int_{\rnum^3 \setminus \Omega} \left(  \mu^{-1} d \VField{e}_{\mathrm{sc}} \wedge d \overline{\VField{v}}  - 
\omega^{2} (\epsilon \VField{e}_{\mathrm{sc}}) \wedge \overline{\VField{v}} \right )
& = & \\
-  \int_{\rnum^{3}\setminus \Omega} \left(  \mu^{-1} d \VField{e}_{\mathrm{inc}} \wedge d \overline{\VField{v}}  - 
\omega^{2} (\epsilon \VField{e}_{\mathrm{inc}}) \wedge \overline{\VField{v}} \right ) & {} & \\
\left( \VField{e}_{\mathrm{int}}-\VField{e}_{\mathrm{inc}} \right)_{|\partial \Omega} & = &  \left( \VField{e}_{\mathrm{sc}} \right )_{|{\partial \Omega}}.
\end{eqnarray*}
Applying a partial integration on the right hand yields
\begin{subequations}
\label{splittedweakequations}
\begin{eqnarray}
\int_{\Omega} \left(  \mu^{-1} d \VField{e}_{\mathrm{int}} \wedge d \overline{\VField{v}}  - 
\omega^{2} (\epsilon \VField{e}_{\mathrm{int}}) \wedge \overline{\VField{v}} \right ) 
+
\int_{\rnum^3 \setminus \Omega} \left(  \mu^{-1} d \VField{e}_{\mathrm{sc}} \wedge d \overline{\VField{v}}  - 
\omega^{2} (\epsilon \VField{e}_{\mathrm{sc}}) \wedge \overline{\VField{v}} \right )
& = & \int_{\partial \Omega} \left(  \mu^{-1} d \VField{e}_{\mathrm{inc}} \wedge \overline{\VField{v}} \right )  \\
\label{splittedweakequationsDirichlet}
\left( \VField{e}_{\mathrm{int}}-\VField{e}_{\mathrm{inc}} \right)_{|\partial \Omega} & = &  \left( \VField{e}_{\mathrm{sc}} \right )_{|{\partial \Omega}}.
\end{eqnarray}
\end{subequations}
On the left hand side we find the quantities of interest we want to compute, namely the total field in the interior $\VField{e}_{\mathrm{int}}$ and the scattered field $\VField{e}_{\mathrm{sc}}$ in the exterior.  The second equation~\eqref{splittedweakequationsDirichlet} merges the scattered and the interior field. After introducing transparent boundary conditions we will explain how to incorporate this field data matching condition~\eqref{splittedweakequationsDirichlet} into a variational formulation

\subsection{Transparent boundary condition (PML)}
\label{sec:pml}
So far, in all considerations the various scattering problems were posed on
the entire domain $\rnum^3$ and are therefore numerically not feasible. This is
overcome by using transparent boundary conditions. We use the perfectly
matched layer method introduced by Berenger \cite{Berenger:94}. This method
exploits the analytic continuation properties of the scattered field in the
exterior domain. In a nutshell using an appropriate complex continuation, the
scattered field is transformed to an exponentially decaying field without
affecting the matching condition with the field in the interior domain. In an
earlier paper we proposed an extremely efficient adaptive PML method which
also copes with non cubic domains and various kinds of inhomogeneous exterior
domains \cite{Zschiedrich2005a, Zschiedrich2006a}. For simplicity, in this paper we restrict
ourselves to cubic domains embedded into an layered media. Let us regard the computational domain $\Omega = [-a,\, a] \times [-b,\,  b] \times [-c,\,  c]$ in Figure~\ref{Fig:IsolatedBlock}. To derive the PML equation we use different coordinate stretchings in each coordinate direction~\cite{Chew:94}, e.g. 
\begin{eqnarray*}
x_\gamma = \left \{ 
\begin{array}{cc}
a+\gamma (x-a), & x>a \\
x, & |x| \leq a\\
-a+\gamma (x+a), & x<-a 
\end{array}
\right ..
\end{eqnarray*}  
The definitions for the $y$ and $z$-directions are accordingly. For a complex coordinate stretching $\gamma$ is a complex number with $\Re{\gamma}=1.0$ and $\Im{\gamma}>0.0$. But firstly we consider $\gamma$ as a real number. Then stretching the coordinates is a simple coordinate transformation. In the differential form calculus, when switching the coordinates, the differential forms (field data) and the operators $\epsilon$ and $\mu$ are transformed accordingly. Subscribing $\gamma$ to the transformed quantities equation~\eqref{splittedweakequations} yields
\begin{eqnarray*}
\int_{\Omega} \left(  \mu^{-1} d \VField{e}_{\mathrm{int}} \wedge d \overline{\VField{v}}  - 
\omega^{2} (\epsilon \VField{e}_{\mathrm{int}}) \wedge \overline{\VField{v}} \right ) 
+
\int_{\rnum^3 \setminus \Omega} \left(  \mu^{-1}_\gamma d \VField{e}_{\mathrm{sc}, \gamma} \wedge d \overline{\VField{v}_\gamma}  - 
\omega^{2} (\epsilon \VField{e}_{\mathrm{sc, \gamma}}) \wedge \overline{\VField{v}_\gamma} \right )
& = & \int_{\partial \Omega} \left(  \mu^{-1} d \VField{e}_{\mathrm{inc}} \wedge \overline{\VField{v}} \right )  \\
\left( \VField{e}_{\mathrm{int}}-\VField{e}_{\mathrm{inc}} \right)_{|\partial
  \Omega} & = &  \left( \VField{e}_{\mathrm{sc}, \gamma} \right )_{|{\partial \Omega}}.
\end{eqnarray*}
This equation is identical to~\eqref{splittedweakequations} but only the transformed quantities are used. Since the test function $\MyField{v}$ is chosen arbitrarily we can avoid using the transformed field and replace $\MyField{v}_\gamma$ by  $\MyField{v}$ without changing the variational form. Now we switch to a complex coordinate stretching. Since the equation above holds true for any real $\gamma$ and due to the holomorphy of the scattered field $\VField{e}_{\mathrm{sc}}$ it is a simple matter of complex function theory that the above equations also hold true for $\gamma$ chosen complex. 
We now want to incorporate the matching condition on the boundary $\partial
\Omega$ into the variational form. At the boundary $\partial \Omega$ the
interior field $\VField{e}_\mathrm{int}$ and the scattered field
$\VField{e}_{\mathrm{sc}, \gamma}$ differ by the field values of the incident field. We therefore add a field with tangential data equal to  $\VField{e}_{\mathrm{inc}}$ on the boundary $\partial \Omega$ and which has local support in the exterior domain $\rnum^3 \setminus \Omega$. Since this field interpolates the data of the incident field on the boundary we denote it by $\mathcal{I} \VField{e}_{\mathrm{inc}}$. In the finite element context $\mathcal{I}$ is the boundary interpolation operator. With the definition $\widetilde{\VField{e}} = \VField{e}_{\mathrm{int}}+\VField{e}_{\mathrm{sc, \gamma}}+\mathcal{I} \VField{e}_{\mathrm{inc}}$ we get the variational formulation
\begin{eqnarray}
\label{weakequationsPML}
\int_{\Omega} \left(  \mu^{-1} d \widetilde{\VField{e}} \wedge d \overline{\VField{v}}  - 
\omega^{2} (\epsilon \widetilde{\VField{e}}) \wedge \overline{\VField{v}} \right ) 
& = & f_{\mathrm{inc}}[\VField{v}] 
\end{eqnarray}
with 
\begin{equation*}
f_{\mathrm{inc}}[\VField{v}]  = -\int_{\Omega_{\mathcal{I}}} \left(  \mu^{-1} d \mathcal{I} \VField{e}_{\mathrm{inc}} \wedge d \overline{\VField{v}}  - \omega^{2} (\epsilon \mathcal{I} \VField{e}_{\mathrm{inc}}) \wedge \overline{\VField{v}} \right ) +
\int_{\partial \Omega} \left(  \mu^{-1} d \VField{e}_{\mathrm{inc}} \wedge \overline{\VField{v}} \right ). 
\end{equation*}
Here, $\Omega_{\mathcal{I}}$ denotes the finite support of $\mathcal{I}
\VField{e}_{\mathrm{inc}}$. Hence on the right hand side only data of the
incident field around the boundary $\partial \Omega$ of the computational
domain are involved. Truncating the exterior domain this variational form is
suitable for a finite element discretization. In the following we again drop
the wiggles.\\
 As we mentioned above deriving the weak form for the system used in the
 domain decomposition iteration~\eqref{strongcoupledsystemupdate} is more
 tedious. In this case interior boundary conditions on the infinite
 hyperplanes $\Gamma_{i, j}$ must be treated. After defining the transparent
 boundary conditions by the PML method these jump conditions are imposed
 within the PML sponge layer. Since the effect of the artificial interior
 boundaries vanishes with convergence of the iterative process the transparent
 boundary conditions on these interfaces only require less accuracy. Even
 approximate transparent boundary conditions can be used.    
\section{Numerical Example: Periodic 3D Mask}
\label{sec:numerical_example}
\begin{figure}
 \psfrag{x}{$x$}
 \psfrag{y}{$y$}
  \begin{center}
      \includegraphics[width=8cm]{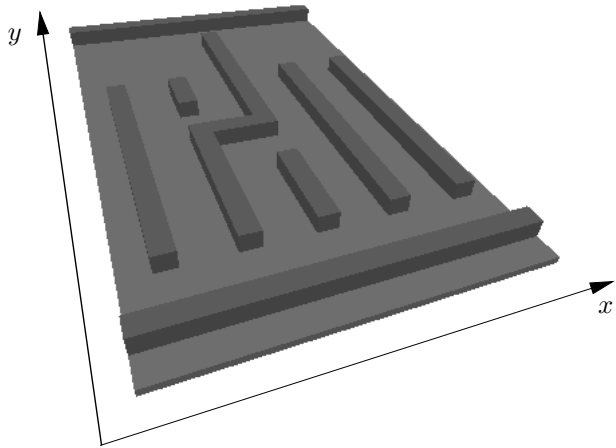}
  \end{center}
  \caption{
\label{Fig:Geometry}
Simulated mask cutout. The mask consists of MoSi-lines with refractive index
$n_{\mathrm{MoSi}} = 2.52+0.596i$ and height $h=65.4nm$. The dimension of the
computational domain is $1.3 \mu \mathrm{m} \times 2.5 \mu \mathrm{m}.$  The
refractive index of the glass substrate is $n_g = 1.5306.$
}
\end{figure}

\begin{figure}
 \psfrag{x}{$x$}
 \psfrag{y}{$y$}
  \begin{center}
      \includegraphics[width=12cm]{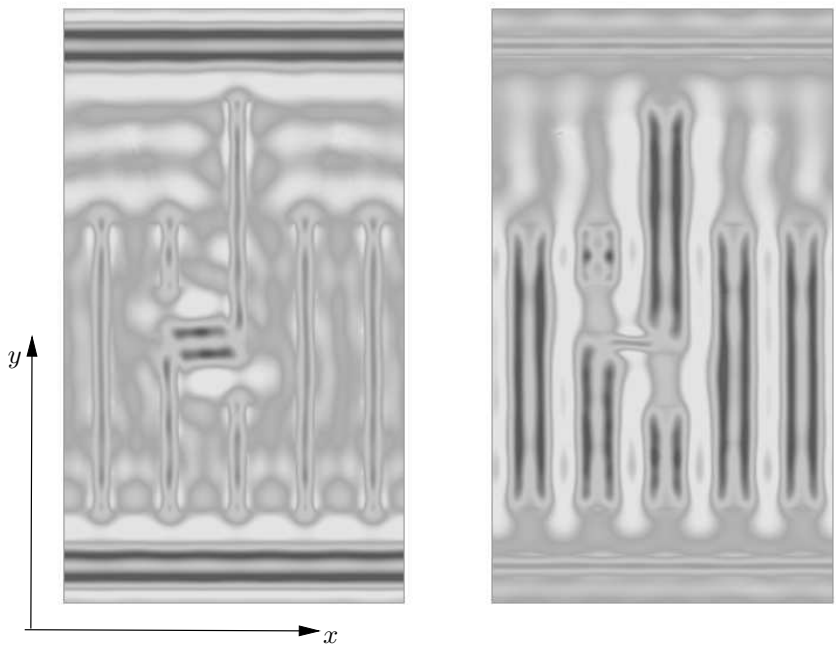}
  \end{center}
  \caption{
\label{Fig:FieldValues}
Near field amplitudes in a distance of $15nm$ above the structure. The incident field is a
plane wave with a vacuum wavelength $\lambda=193 \mathrm{nm}.$ For the left
pseudo-color plot the incident field is $x$-polarized and for the right one
the incident field is $y-polarized.$
}
\end{figure}

In this section we apply the domain decomposition method to a 3D mask cutout
given in Figure~\ref{Fig:Geometry}. The structure consists of MoSi-lines of
height $h=65.4 \mathrm{nm}$ on a glass substrate. Here, the sidewall angle of
the lines is $90 \mathrm{deg}$, but other sidewall angle parameter can be treated
without extra costs. Further material values are given in the description to Figure~\ref{Fig:Geometry}.
We simulated the mask with high order finite elements of order 5 and 6. 
Using finite element degree 5 (1.6 million unknowns) it was still possible to
compute the solution with the direct method PARDISO~\cite{PARDISO}. A
comparison of the domain decomposition method and the direct method shows a
good agreement. After five iterations the domain decomposition method with 16
vertical sub-domains converged to the exact solution up to an error of
$10^{-5}$ in the complex amplitudes of the  diffraction modes. The required
memory was reduced by a factor of 3 (from 71GB to 25GB). On our computer it
was not possible to compute the solution to finite element degree 6 by a
direct sparse LU method due to limited memory resources. However, the
convergence rate of the domain decomposition method was not affected when
increasing the finite element degree from 5 to 6. This way we were able to
solve the linear system with more than 3.3 million of unknowns. However with
50GB needed RAM the memory requirements are still demanding. To reduce further the memory demand it seems promising to split the domains also in the horizontal direction, which will be done in future work. In Figure~\ref{Fig:FieldValues} the computed near fields for an perpendicular incidence of $x-$ and $y-$ polarized light is plotted in a distance of $15nm$ above the structure.
\section{Conclusions}
We have proposed a new rigorous domain decomposition method for Maxwell's
equations. This method allows for the reduction of needed computer
resources. Especially it is possible to reduce significantly the amount of
needed memory. Is has been shown that the method converges to the exact
(discrete) finite element solution which corresponds to a large scale linear
system of equations. The convergence rate behaviour is currently under investigation.

\bibliography{lit}     
\bibliographystyle{spiebib}   
\end{document}

%% file: MatheDef.tex
\newcommand{\rnum}{{\bf {R}}}

\newcommand{\Alt}{{\mathrm{Alt}}}

\newcommand{\VField}[1]{{\bf #1}} 
\newcommand{\MyField}[1]{{\bf #1}} 
\newcommand{\ltl}{L^2_{\rm loc}}
\newcommand{\hcl}{H_{\rm loc}({\rm curl})}

\def\squareforqed{\hbox{\rlap{$\sqcap$}$\sqcup$}}
\def\qed{\ifmmode\else\unskip\quad\fi\squareforqed}